# High-field superconductivity in alloyed MgB$_2$ thin films

V. Braccini[1], A. Gurevich[1], J.E. Giencke[1], M.C. Jewell[1], C.B. Eom[1], D.C. Larbalestier[1], A. Pogrebnyakov[2], Y. Cui[2], B. T. Liu[2], Y. F.Hu[2], J. M. Redwing[2], Qi Li[2], X.X. Xi[2], R.K. Singh[3], R. Gandikota[3], J. Kim[3], B. Wilkens[3], N. Newman[3], J. Rowell[3], B. Moeckly[4], V. Ferrando[5], C. Tarantini[5], D. Marré[5], M. Putti[5], C. Ferdeghini[5], R. Vaglio[6], E. Haanappel[7]

[1]Applied Superconductivity Center, University of Wisconsin – Madison, WI

[2]Pennsylvania State University, University Park, PA

[3]Arizona State University, Tempe, AZ

[4]Superconductor Technologies Inc, Sunnyvale, CA

[5]University of Genoa / INFM-LAMIA, Genoa, Italy

[6]University of Naples / INFM-Coherentia, Naples, Italy

[7]Laboratoire National des Champs Magnétiques Pulses, CNRS-UPS-INSA, Toulouse, France

We investigated the effect of alloying on the upper critical field H$_{c2}$ in 12 MgB$_2$ films, in which disorder was introduced by growth, carbon doping or He-ion irradiation, finding a significant H$_{c2}$ enhancement in C-alloyed films, and an anomalous upward curvature of H$_{c2}$(T). Record high values of H$_{c2}^{\perp}$(4.2) ≈ 35T and H$_{c2}^{\parallel}$(4.2) ≈ 51T were observed perpendicular and parallel to the *ab* plane, respectively. The temperature dependence of H$_{c2}$(T) is described well by a theory of dirty two-gap superconductivity. Extrapolation of the experimental data to T=0 suggests that H$_{c2}^{\parallel}$(0) approaches the paramagnetic limit of ~70T.

Discovery of superconductivity in $MgB_2$ with the critical temperature $T_c = 39K$ renewed intense interest in the novel effects in two-gap superconductors. *Ab initio* calculations[1,2] showed that $MgB_2$ has two weakly coupled gaps $\Delta_\sigma(0) \approx 7.2meV$ and $\Delta_\pi(0) \approx 2.3meV$ residing on disconnected sheets of the Fermi surface formed by in-plane $p_{xy}$ boron orbitals ($\sigma$ band) and out-of-plane $p_z$ boron orbitals ($\pi$ band). The two-gap Eliashberg theory[2,3] has explained many anomalies in tunneling, heat capacity and electrodynamics of clean $MgB_2$ single crystals.[4] However, the physics of two-gap $MgB_2$ alloys determined by the multiple impurity scattering channels, and by the complex substitutional chemistry of $MgB_2$[5] are still poorly understood. The behavior of disordered $MgB_2$ is particularly interesting because it exhibits enormous enhancement of $H_{c2}$ by nonmagnetic impurities,[6-8] well above the estimate $H_{c2}(0)=0.69T_cH'_{c2}(T_c)$ of one-gap theory[9], and anomalous temperature-dependent $H_{c2}$ anisotropy.[4] Some of these features have been recently explained by generalized two-gap Usadel equations[10,11] in which impurity scattering is accounted for by the intraband electron diffusivities $D_\sigma$ and $D_\pi$, and interband scattering rates $\Gamma_{\sigma\pi}$ and $\Gamma_{\pi\sigma}$. In this Letter we present high-field transport measurements of $H_{c2}(T)$ for 12 $MgB_2$ films made by 6 different groups using very different ways of introducing disorder. We show that $H_{c2}$ is radically increased in dirty films, and $H_{c2}^{\parallel}(0)$ extrapolates to $H_p \sim$ 70T for a C-alloyed film, comparable to the paramagnetic limit ($H_p=1.84T_c = 64T$ for $T_c = 35K$).

Our films were made by different deposition techniques including pulsed laser deposition (PLD),[12,13] molecular beam epitaxy (MBE),[14] hybrid physical-chemical vapor deposition (HPCVD),[15] sputtering,[16,17], and reactive evaporation.[18] Growth was performed by *in-situ* techniques[14,15,17,18], as well as *ex-situ* methods with post-annealing in Mg vapor[12,13]. C-doped films were produced by HPCVD[15] with the addition of 75 sccm of $(C_5H_5)_2Mg$ to the $H_2$ carrier gas. Some films were damaged with $10^{16}$ cm$^{-2}$, 2 MeV $\alpha$-particles to controllably alter the scattering by irradiation point defects.[19] Elemental compositions were determined by wavelength dispersive spectroscopy (WDS) and Rutherford Backscattering Spectroscopy (RBS), and film orientation and lattice parameters with a four-circle X-ray diffractometer. Film parameters are summarized in Table I. In some samples RBS detected through-thickness composition variations, likely due to surface reactions

Measurements of $H_{c2}(T)$ on samples A, B, E, F, H, I, L were performed in a 33T resistive magnet at the NHMFL in Tallahassee. Film resistances R(H) were measured in parallel and perpendicular fields at a sweep rate of 1T/min while temperature was stabilized to ~10mK. The measuring current density J was varied between 10 and 100A/cm$^2$. Detailed study of film A did not show any significant change in R(H) for 4 < J < 4000 A/cm$^2$. Samples G, M and N were measured in the 300 msec 60 T pulsed facility at the LNCMP in Toulouse, at a lock-in frequency of 40 KHz and J varying from 50 to 200 A/cm$^2$ with no change in R(H). In all cases $H_{c2}$ was defined as $R(H_{c2}) = 0.9R(T_c)$.

Figure 1 shows $H_{c2}^\perp(T)$ (a) and $H_{c2}^\parallel(T)$ (b) for the lower $H_{c2}$ samples A, B, C, E, H, I, L, M, N, for which a wide variety of properties can be developed. R(H) curves for film A are shown in inset. The $H_{c2}^\perp(T)$ data in Fig. 1a fall into two groups, one having $T_c\approx$32-37K, with relatively low $dH_{c2}/dT$ and $H_{c2}(0)$ ~10.5-15T, while the lower $T_c$ group (24–32K) has ≈ 50% larger $dH_{c2}/dT$ and $H_{c2}(0)$~17–22T. $H_{c2}(0)^\parallel$ data in Fig. 1b range more continuously from 18-40T, with only samples B and L standing out. Film B, with the lowest $T_c\approx$24K and $H_{c2}(0)$ with $\rho_n$~85μΩcm, has almost identical $H_{c2}^\parallel(T)$ and $H_{c2}^\perp(T)$, while non-textured sample L with $\rho_n$~9.9μΩcm also has a low $H_{c2}(0)$~22T in spite of its higher $T_c$=39.4K. Film E, with highest $T_c$=41.5 K and $\rho_n$~0.4μΩcm as made (1.6μΩcm when measured at the NHMFL) represents MgB$_2$ in the clean limit[20], making it unsuitable for fitting using dirty-limit theory.[8] Although film E has the lowest $\rho_n$, it exhibits the highest $H_{c2}^\parallel(T)$, even though Fig. 1b also includes films with $\rho_n$>500μΩcm but with lower $H_{c2}^\parallel$ (film I). Thus, there is no simple correlation between $\rho_n$ and $H_{c2}$, because the global resistivity may be limited by poor intergrain connectivity[21] while $H_{c2}$ is controlled by intragrain impurity scattering. The anisotropy parameter $\gamma(T)=H_{c2}^\parallel/H_{c2}^\perp$ ranges from ≈3 for the lowest $\rho_n$ film (E, $T_c$=41.5K) to ≈1 for the lowest $T_c$ textured film B ($T_c$=24K). For most films, $\gamma(T)$ tends to decrease as T decreases, consistent with the behavior predicted for two-gap MgB$_2$ with dirtier π band [10].

Figure 2 shows $H_{c2}^\parallel(T)$ and $H_{c2}^\perp(T)$ curves for the highest $H_{c2}$ films D, F and G, while the insets show the parallel-field R(H) traces. By increasing the nominal carbon content in the HPCVD films, resistivity rises from ~1.6 (E) to 564 (F) and 250 μΩcm (G), while $T_c$ only decreases to 35K. However, $H_{c2}^\perp(0)$ increases from 12T (E) to 28T (G) and ≈ 40T (F). Furthermore, $H_{c2}^\parallel(0)$ rises from ≈ 35T (E) to 51T (G) and more than 70T in sample F, while the anisotropy parameter $\gamma(T)=H_{c2}^\parallel(T)/H_{c2}^\perp(T)$ decreases as $\rho_n$ increases. Figure 2c presents $H_{c2}(T)$ for sample D, which has high nominal O (17at.%) and C (14at.%) content.[8] Unlike the two *in situ* films made by HPCVD, film D was made *ex situ* by PLD. This film has $H_{c2}^\perp(0)\approx$33T and $H_{c2}^\parallel(0)\approx$48T.

The $H_{c2}(T)$ curves in Fig. 2 have an upward curvature inconsistent with the dirty limit one-band theory.[9] For two-gap pairing, intraband scattering does not affect $T_c$, but $T_c$ decreases as the pair-breaking interband scattering parameter $g = (\Gamma_{\sigma\pi}+\Gamma_{\pi\sigma})\hbar/2\pi k_B T_{c0}$ increases, where $T_{c0} = T_c(g=0)$. Due to the orthogonality of the $\sigma$ and $\pi$ orbitals in MgB$_2$, g is usually small, and $T_c$ does not change much, even if $\rho_n$ is significantly increased.[25] The insensitivity of $T_c$ to scattering makes it possible to increase $H_{c2}$ in MgB$_2$ to a much greater extent than in one-gap superconductors by optimizing the diffusivity ratio $D_\pi/D_\sigma$. The equation for $H_{c2}$ and $T_c$ in a dirty two-gap superconductor has the form[10]

$$2w(\ln t + U_+)(\ln t + U_-) + (\lambda_0 + \lambda_i)(\ln t + U_+) + (\lambda_0 - \lambda_i)(\ln t + U_-) = 0, \qquad (1)$$

$$\psi\left(\frac{1}{2}+\frac{g}{t_c}\right) - \psi\left(\frac{1}{2}\right) = -\frac{2\ln t_c (w \ln t_c + \lambda_0)}{2w \ln t_c + \lambda_0 + \lambda_i}, \qquad (2)$$

where $t = T/T_{c0}$, $t_c = T_c/T_{c0}$, $T_{c0}=1.13\theta_D \exp[-(\lambda_-+\lambda_0)/2w]$, $w=\lambda_{\sigma\sigma}\lambda_{\pi\pi}-\lambda_{\sigma\pi}\lambda_{\pi\sigma}$, $\lambda_0=(\lambda_-^2+4\lambda_{\sigma\pi}\lambda_{\pi\sigma})^{1/2}$, $\lambda_\pm=\lambda_{\sigma\sigma}\pm\lambda_{\pi\pi}$, $\lambda_{mn}$ is 2×2 matrix of BCS coupling constants, $\lambda_{\sigma\sigma}$ and $\lambda_{\pi\pi}$ describing intraband pairing, and $\lambda_{\sigma\pi}$ and $\lambda_{\pi\sigma}$ describing interband pairing. Here $\lambda_i=[(\omega_-+\Gamma_-)\lambda_- -2\lambda_{\sigma\pi}\Gamma_{\pi\sigma}-2\lambda_{\pi\sigma}\Gamma_{\sigma\pi}]/\Omega_0$, $\Gamma_\pm =\Gamma_{\sigma\pi} \pm \Gamma_{\pi\sigma}$, $\omega_\pm = (D_\sigma\pm D_\pi)\pi H/\phi_0$, $\Omega_0=(\omega_-^2+\Gamma_+^2+2\Gamma_-\omega_-)^{1/2}$, $U_\pm(H,T) = \psi(1/2+\hbar\Omega_\pm/2\pi k_B T) - \psi(1/2)$, $\Omega_\pm = \omega_+ + \Gamma_+ \pm \Omega_0$, where $\psi(x)$ is the digamma function, and $\phi_0$ is the flux quantum. For **H**||ab, the diffusivities in Eq. (1) should be replaced according to: $D \rightarrow [D^{(ab)}D^{(c)}]^{1/2}$.[10]

The evolution of $H_{c2}(T)$ and $T_c$ with g is shown in Fig. 3. For dirty $\pi$ band ($D_\pi \ll D_\sigma$) and g = 0, $H_{c2}(T)$ has an upward curvature, because $dH_{c2}/dT$ at $T_c$ is determined by the larger $D_\sigma$, while $H_{c2}(0)$ is determined by the smaller $D_\pi$. As g increases, the upward curvature of $H_{c2}(T)$ diminishes, and $T_c$ decreases. In the limit $g \gg 1$ of complete interband mixing, $T_c(g)$ approaches $T_{min}=T_{c0}\exp[-(\lambda_0+\lambda_i)/2w]$, while $H_{c2}(T)$ for fixed $D_\pi$ and $D_\sigma$ evolves toward the one-gap $H_{c2}(T)$ curve. The case of a dirtier $\sigma$ band, $D_\pi \gg D_\sigma$ corresponds to a one-gap-like $H_{c2}(T)$ curve broadened near $T_c$ by the band mixing.

We used Eqs. (1)-(2) to describe the observed $H_{c2}(T)$, taking *ab-initio* values[3] $\lambda_{\sigma\sigma}=0.81$, $\lambda_{\pi\pi}=0.28$, $\lambda_{\sigma\pi}=0.115$ and $\lambda_{\pi\sigma}=0.09$ as input parameters. First, g was calculated from Eq. (2) with the observed $T_c$ and $T_{c0} = 39$K. Next, we calculated $D_\sigma$ from the observed (or extrapolated) $H_{c2}(0)$, leaving the ratio $D_\pi/D_\sigma$ as the only fit parameter determining the shape of $H_{c2}(T)$. This procedure is based on a conventional assumption of the dirty limit theory that impurities only change the scattering rates, but do not affect the coupling constants $\lambda_{mn}$, or the electron density of states. The fits describe well the observed $H_{c2}(T)$ curves in Fig. 2, indicating that $\pi$ scattering is stronger than $\sigma$ scattering in all our high-$H_{c2}$ films. The extrapolated $H_{c2}^{\parallel}(0)$ reaches $\approx$ 55T for film G and >70T for film F.[26]

Our $H_{c2}(T)$ data are striking because the highest $H_{c2}$ values are attained for films with weak $T_c$ suppressions, and the three highest-$H_{c2}$ films ($48 < H_{c2}^{\parallel}(0) < 70T$) greatly exceed the $H_{c2}^{\parallel}(0)$ values reported for C-doped $MgB_2$ single crystals (~35T)[22,23] and C-doped filaments (32T).[24] The fits to $H_{c2}(T)$ for films D, F, and G, which all have significant C content, indicate much stronger $\pi$ than $\sigma$ scattering.

We find that the very broad range of $\rho_n$ (~1-600$\mu\Omega$cm) does not directly manifest itself in the atomic-scale scattering that actually determines $H_{c2}$, because current-blocking extended defects[21] control the measured resistivity. TEM on a C-doped HPCVD film found significant amorphous C-rich phase and scattered $MgC_2$ precipitates surround many $Mg(B_{1-x}C_x)_2$ grains, thus showing that much less than the total C content from Table I is dissolved in $MgB_2$. Because $T_c$ is depressed to zero at x ~ 0.15 ($\approx$10 at%) and $\rho_n \geq$ 50$\mu\Omega$cm in $Mg(B_{1-x}C_x)_2$ single crystals[22,23] and filaments[24], this would indicate that films with $T_c$ of 33-35K have x~0.03-0.05 within the $MgB_2$ grains. This heterogeneous microstructure results in $T_c$ inhomogeneities, which manifest themselves in the resistive transition broadening in Figs. 2.

To understand the scattering mechanisms better, we observe that, similar to C-alloyed $MgB_2$ single crystals,[28] our high-$H_{c2}$ films F and G have smaller a-axis lattice spacing than the clean limit value of 0.3085nm. However, the *c*-axis lattice spacing in our high-$H_{c2}$ films is larger than the value of 0.3524nm measured for bulk pure and C-alloyed single crystals and filaments.[24,28] TEM study of film D showed buckling of the *ab* planes[8] (perhaps due to strains induced by as-grown nanophase precipitates), naturally causing the c-axis lattice expansion. Furthermore, lattice buckling results in strong $\pi$ scattering due to disturbance of the $p_z$ $\pi$ orbitals, and thus dirtier $\pi$ band ($D_\pi \ll D_\sigma$) necessary to account for the upward curvature of $H_{c2}(T)$ in Figs. 2b and 2c. This scenario may also explain how C (which normally substitutes for B) can result both in the strong in-plane $\sigma$ band scattering and out-of-plane $\pi$ scattering required for the observed $H_{c2}$ enhancement.

In conclusion, we have performed extensive studies of the effect of disorder on $H_{c2}$ of $MgB_2$ and report record high $H_{c2}$ values, which may approach the paramagnetic limit for C-doped films. The upward curvature of $H_{c2}(T)$ and weak $T_c$ suppression are described well by the two-gap theory.

Work at UW was supported by NSF under MRSEC grant DMR-0079983, work at Penn State was supported under ONR Nos. N00014-00-1-0294 (Xi) and N0014-01-1-0006 (Redwing), by NSF under grant Nos. DMR-0306746 (Xi and Redwing), DMR-9876266 and DMR-9972973 (Li), and work at Arizona State U was supported under ONR Nos. N00014-02-1-0002. The NHMFL user facility is supported by NSF.

**Figure captions**

**Figure 1:**

$H_{c2}^{\perp}(T)$ (a) and $H_{c2}^{\parallel}(T)$ (b) for samples A, B, C, E, H, I, L, M, and N. The lines are guides for the eye. Insets show R(H) for sample A for: T = 2.1, 4.2, 8, 10, 15, 20, 22, 25, 30 K (a), and T = 2.47, 3.34, 4.2, 6.8, 10, 12, 15, 18.2, 20, 22, 25, 28, 32 K (b). The $R(H_{c2}) = 0.9R_n$ criterion used to determine $H_{c2}$ is shown with a dashed line.

**Figure 2:**

$H_{c2}^{\parallel}(T)$ (triangles) and $H_{c2}^{\perp}(T)$ (squares) for films G (a), F (b) and D (c). Insets show the raw R(H) traces for H||*ab*. Solid curves are calculated from Eqs. (1) and (2) with fit parameters given in Table I.

**Figure 3:**

$H_{c2}(T)$ curves calculated from Eq. (1) for $D_\pi = 0.03 D_\sigma$ and g = 0.01, 0.05, 0.2, 1, 10 (from top to bottom). Inset shows $T_c(g)$ calculated from Eq. (2) with $\lambda_{mn}$ taken from Ref. 3.

**Table I**

| Samples | Substrate | $T_c$ (K) | $\rho_n$(40K) ($\mu\Omega$cm) | $H_{c2}^\perp$ (T) | $H_{c2}^\parallel$ (T) | g | $D_\pi/D_\sigma$ | c (Å) | a (Å) | Mg at% | B at% | C at% | O at% |
|---|---|---|---|---|---|---|---|---|---|---|---|---|---|
| A epitaxial [16] | (0001)Al$_2$O$_3$ | 35 | 9(4) | 13.5 | 33 | 0.045 | 0.12 | 3.516 | 3.047 | 29 | 53 | 10 | 8 |
| B fiber-textured [16] | (0001)Al$_2$O$_3$ | 23.7 | 86(56) | 17 | 17 | 0.5 | <<1 | - | - | 28 | 57 | 7 | 8 |
| C epitaxial*[16] | (0001)Al$_2$O$_3$ | 34 | 7 | 20.5 | 30 | 0.06 | <<1 | 3.52 | 3.08 | - | - | - | - |
| D fiber-textured*[13] | (111)SrTiO$_3$ | 31 | 220 | 33 | 48 | 0.075 | <<1 | 3.547 | - | 37 | 32 | 14 | 17 |
| E epitaxial | SiC | 41.5 | 1.6(0.4) | 12 | 34.5 | - | - | 3.511 | 3.107 | 30 | 57 | 2 | 11 |
| F fiber-textured [15] | SiC | 35 | 564 | 40 | >74 | 0.045 | <<1 | 3.542 | 3.055 | 26 | 46 | 21 | 6 |
| G fiber-textured [15] | SiC | 35 | 250 | 28.2 | 55.5 | 0.045 | 0.065 | 3.536 | 3.057 | 25 | 42 | 26 | 6 |
| H epitaxial [15] | SiC | 38 | 10.5 | 10.5 | 30 | 0.025 | 0.06 | 3.519 | 3.107 | 31 | 63 | 4 | 1 |
| I untextured [14] | (0001)Al$_2$O$_3$ | 32 | 567(290) | 21.7 | 26.8 | 0.09 | 0.08 | - | - | - | - | - | - |
| L no 00l textured [18] | r-cut Al$_2$O$_3$ | 39.4 | 9.9(2.8) | 10.8 | 21.4 | 0.025 | 0.07 | - | - | 32 | 65 | 1 | 1 |
| M epitaxial [12] | (111)MgO | 33.5 | 47 | 14.6 | 38.1 | 0.095 | 0.1 | 3.533 | 3.036 | 24 | 41 | 28 | 6 |
| N untextured [17] | (001)MgO | 28.6 | 400 | 15.8 | 24.3 | 0.155 | <<1 | - | - | 33 | 53 | 5 | 9 |

Sample list with substrate, texture and lattice parameters derived from XRD and chemical compositions deduced from WDS. Impurities detected in amounts less than 1 at.% are not listed. $\rho_n$(40K) was obtained during $H_{c2}$ measurements (as-grown values are given in parentheses). $H_{c2}^\parallel$ and $H_{c2}^\perp$ values were extrapolated to 0K, and g and $D_\pi/D_\sigma$ were deduced from the fit of $H_{c2}(T)$ curves for all films. ($D_\pi/D_\sigma$ << 1 means that the data point scatter does not allow us to distinguish between finite and zero $D_\pi/D_\sigma$, so the fit was performed for $D_\pi = 0$).

Figure 1 a

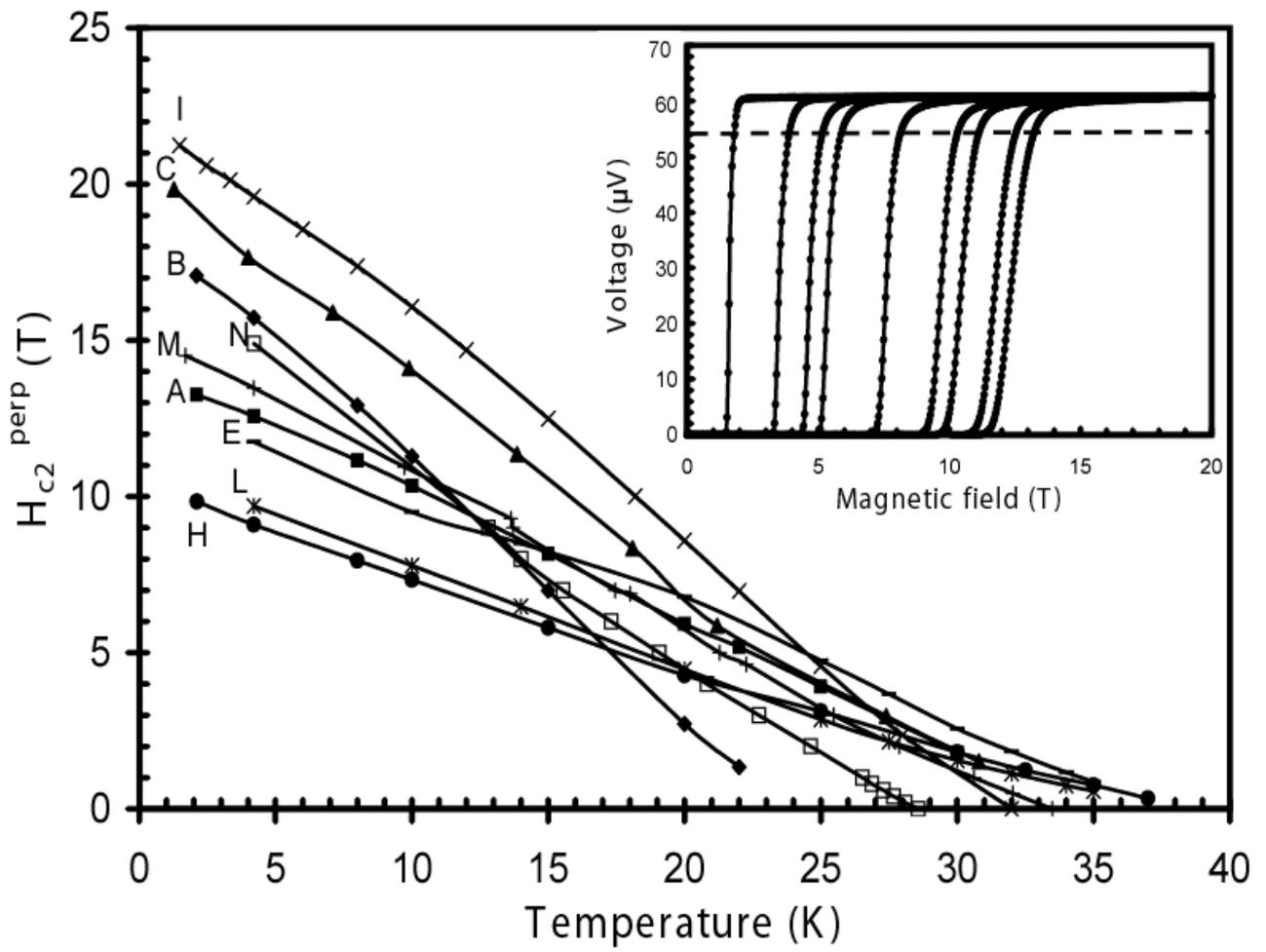

Figure 1 b

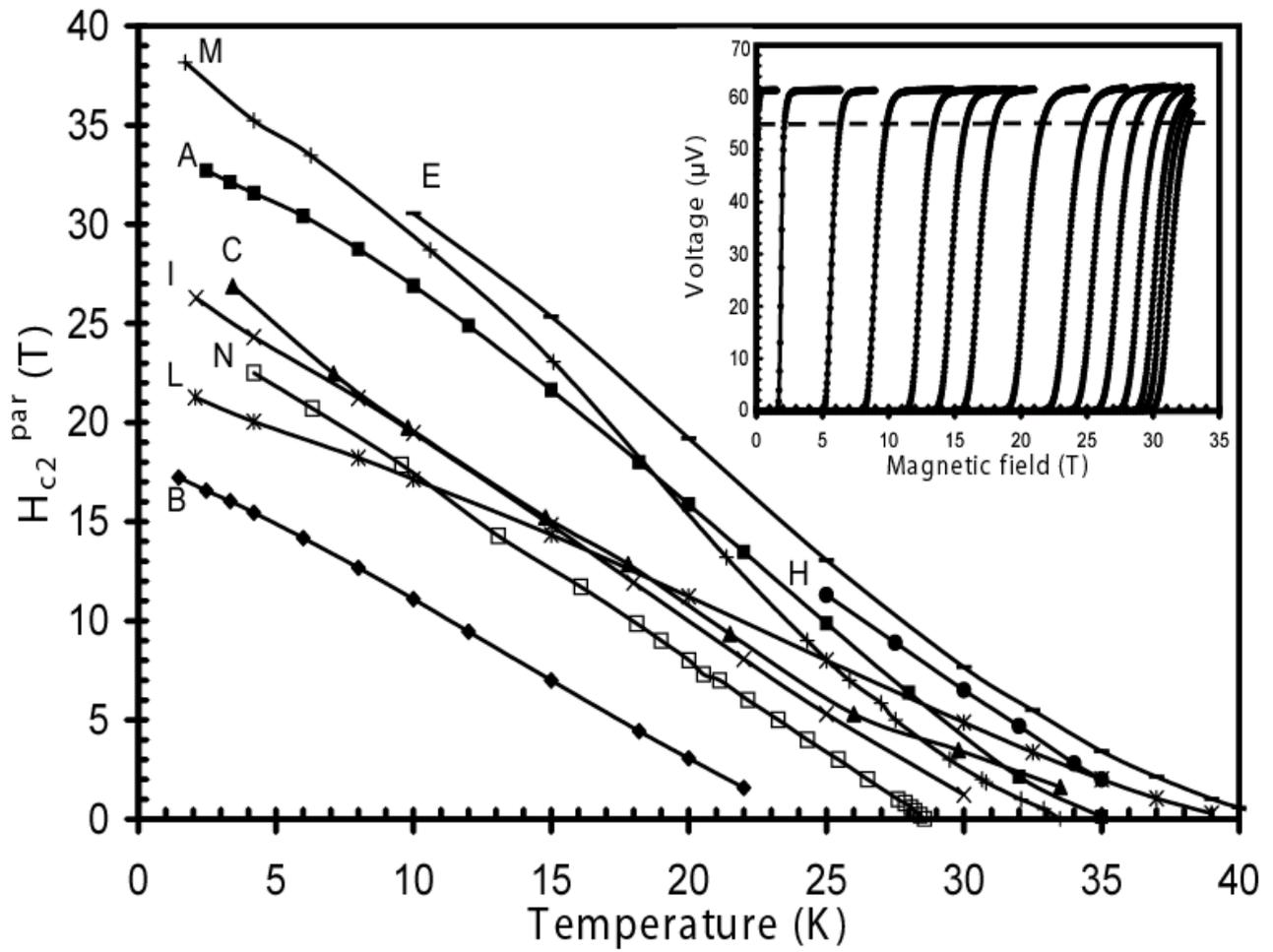

Figure 2 a

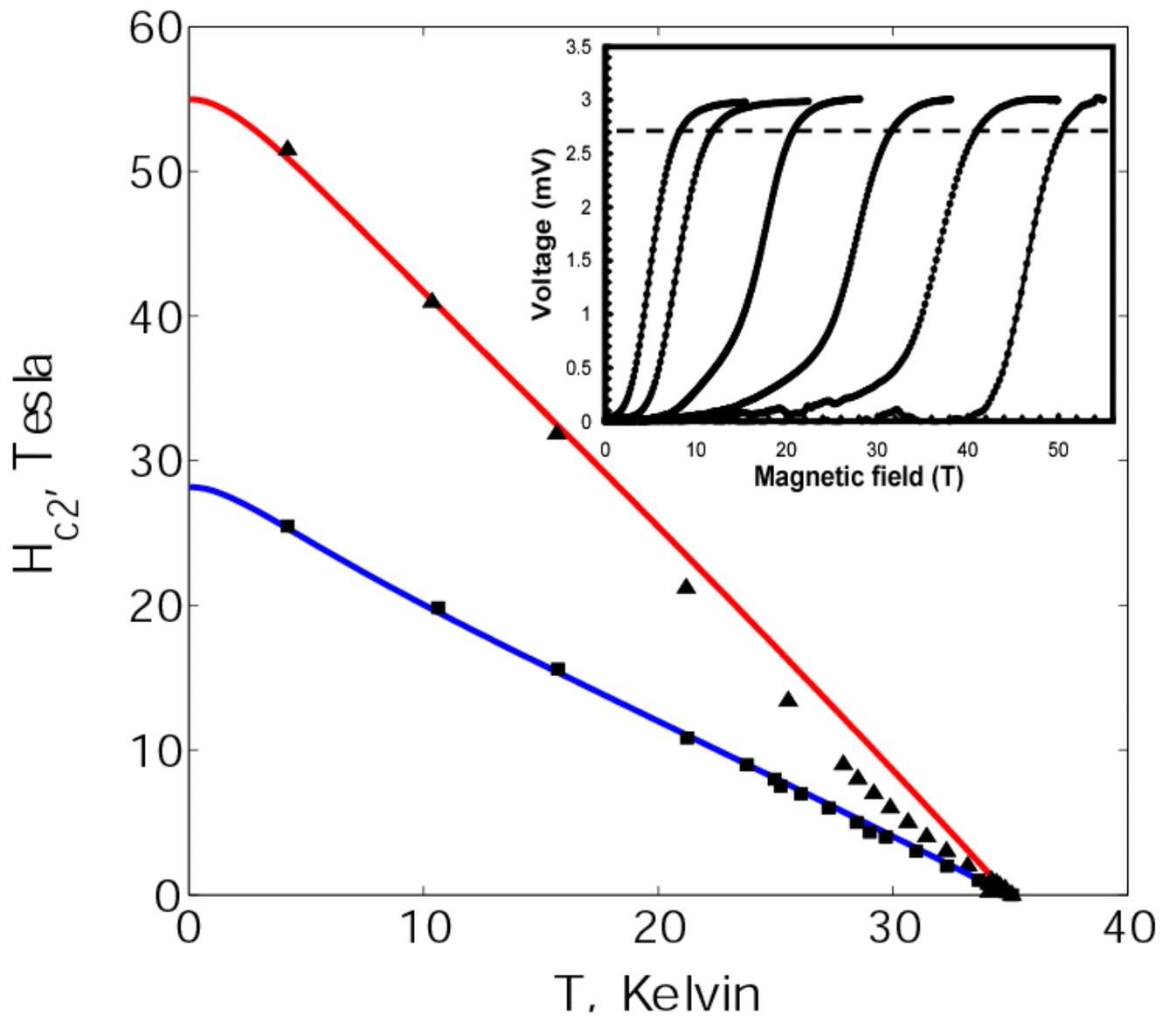

Figure 2 b

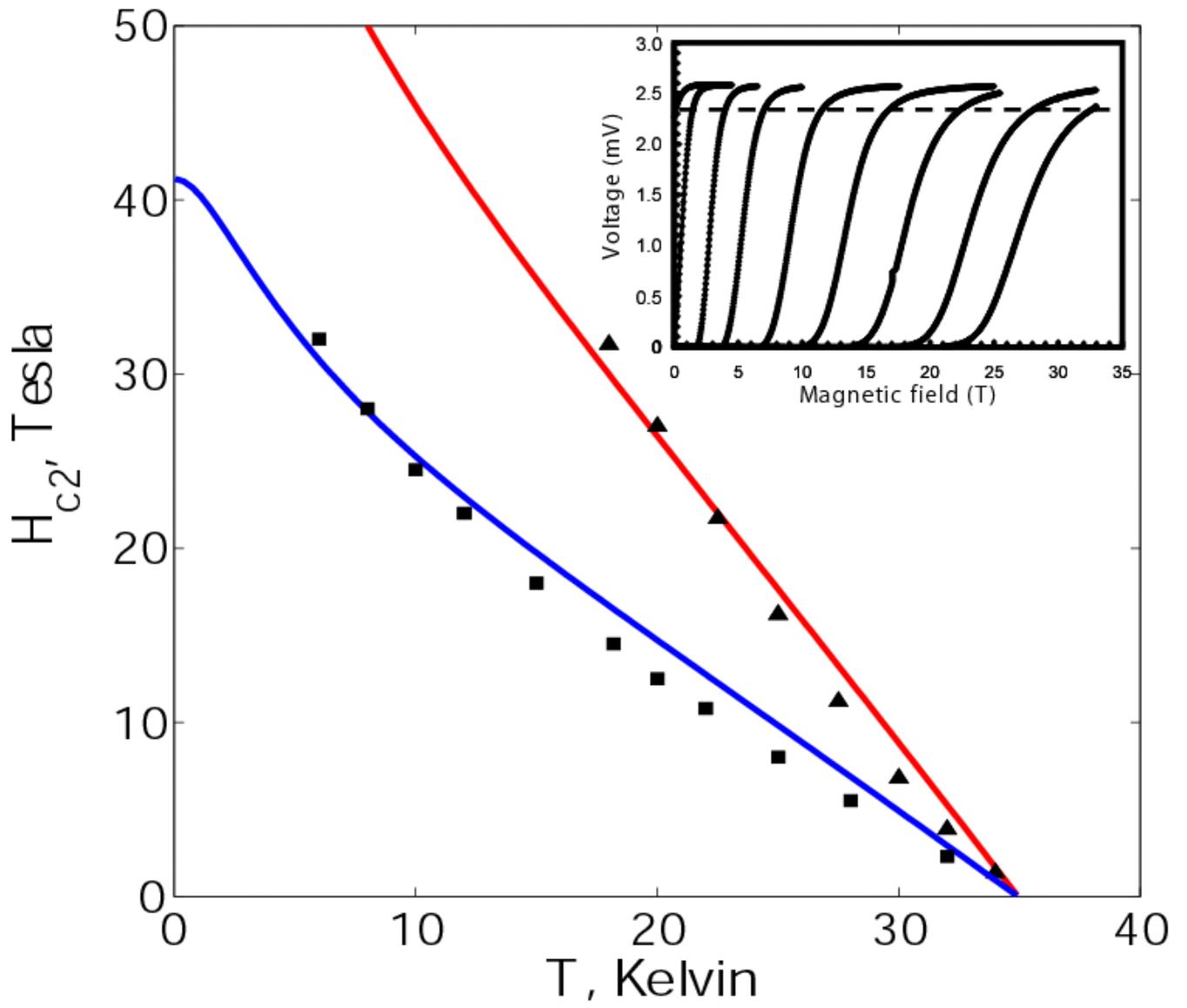

Figure 2 c

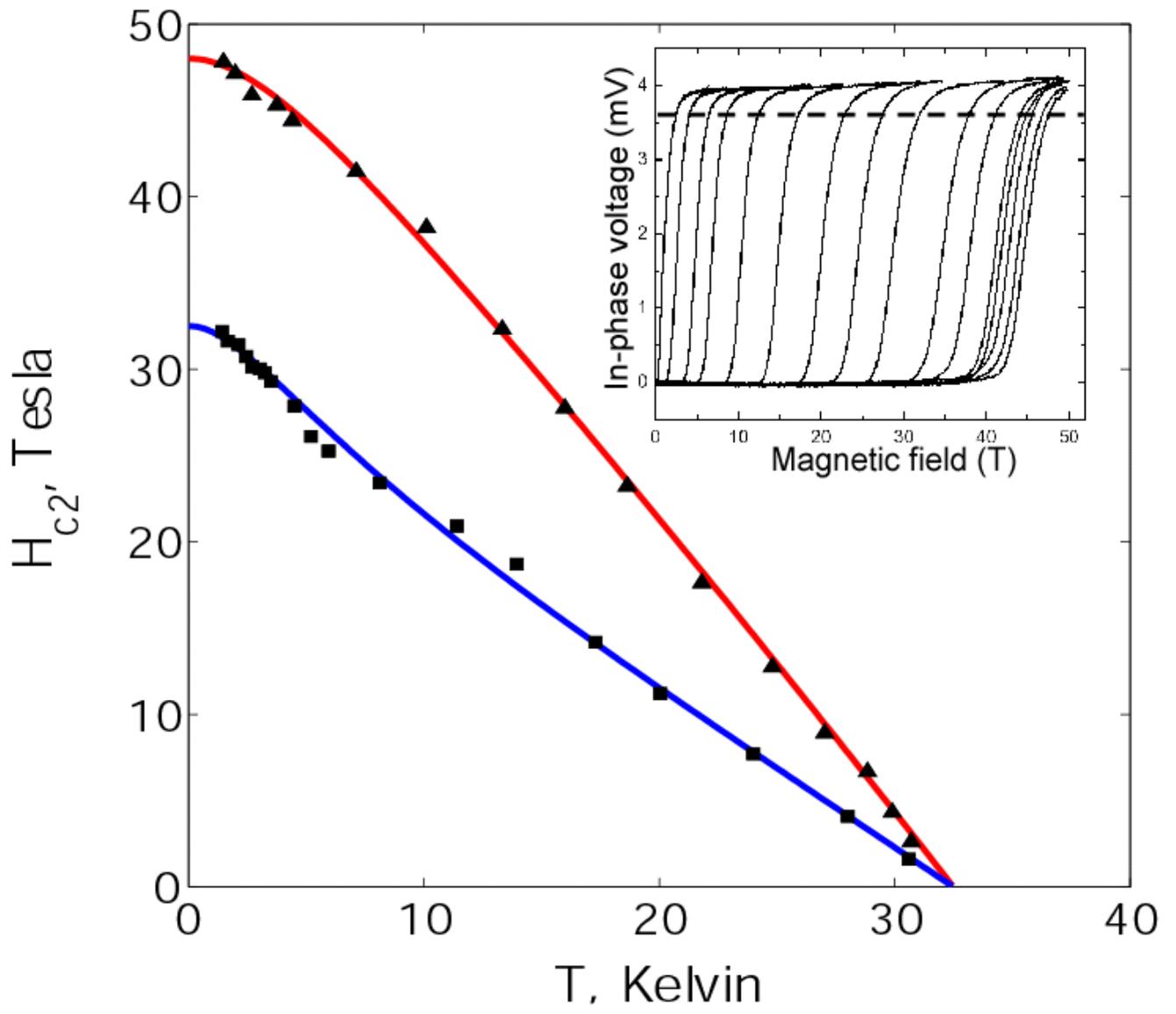

Figure 3

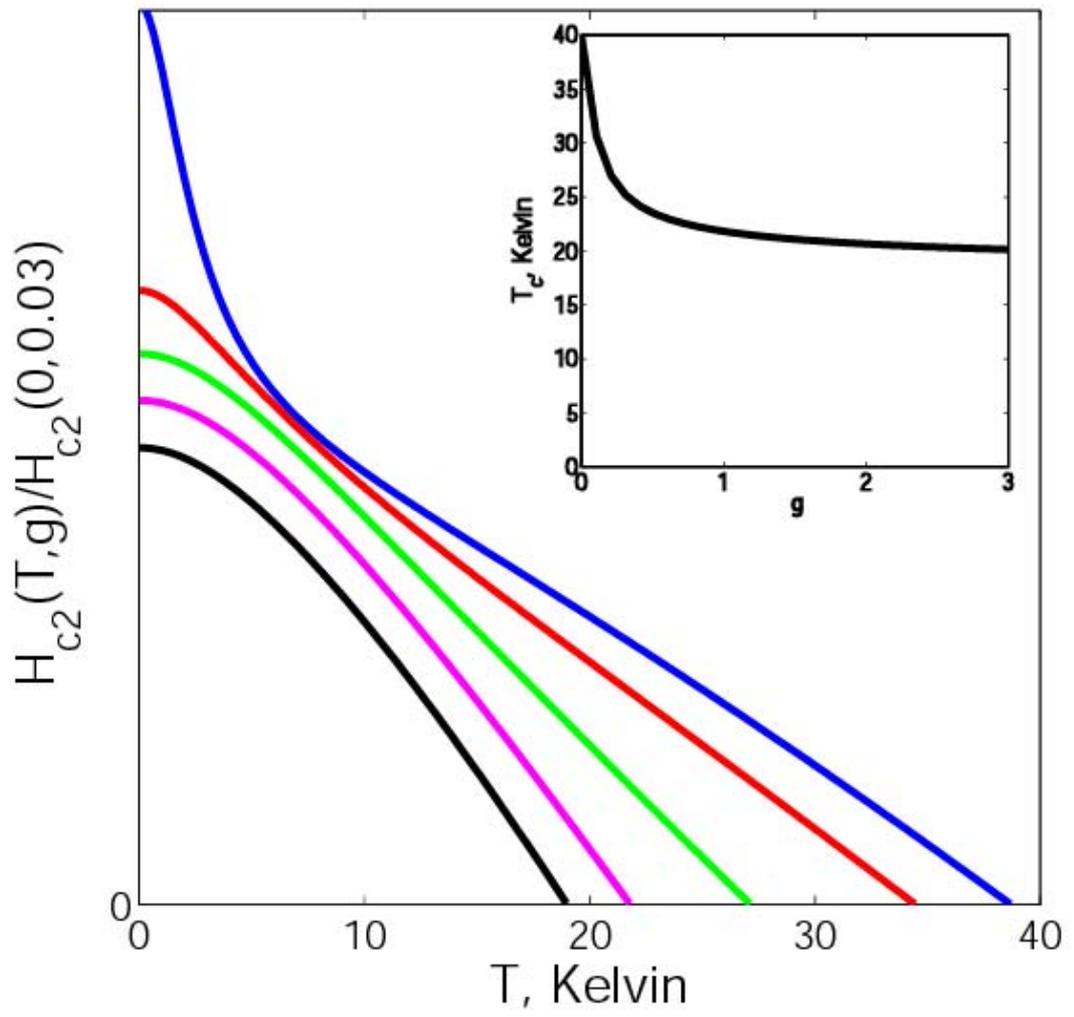